\documentclass[12pt,a4paper]{article}
\usepackage{epsf}
\newcommand{\nav}{\left[\left\langle n \right\rangle\right]}
\begin{document}
\title{Large alphabet quantum key distribution with two-mode coherently correlated beams}
\author{Vladyslav C. Usenko\dag 
\footnote{To whom correspondence should be addressed (usenko@iop.kiev.ua)} and Bohdan I. Lev\dag \\\\
\dag\ \it Institute of Physics, National Academy of Sciences of Ukraine, \\ \it 46 Nauky pr, Kyiv 03028, Ukraine}

\maketitle

\begin{center}

\end{center}

\begin{abstract}
The large-alphabet quantum cryptography protocol based on the two-mode coherently correlated multi-photon beams is proposed. The alphabet extension for the protocol is shown to result in the increase of the QKD effectiveness and security.
\end{abstract}


\section{Introduction}
Quantum cryptography \cite{main}, being the first practical realization of quantum physics at the single quanta level, is the art of creating data channels physically secure against eavesdropping. Most of its successful practical realizations are 
based on the use of single photon information coding though optical fiber links thus being based on a weak laser pulses. The security of the protocols is based on the state perturbation during the eavesdropping \cite{bb84} or the measurement correlations analysis with the Bell inequalities check \cite{e91}. Since key bits are encoded in the single photon states the appearance of the additional photons may seriously undermine the protocol security by leading to the successful listening-in because additional photons' states can be imperceptibly measured by an eavesdropper. Thus the security reasons require the minimizing of the multi-photon pulses appearance probability making most of the pulses empty, which limits the key rate and results in an additional error rate caused by the single-photon counting detectors, which are inclined to "dark counts", clicking when the photon is missing \cite{main}.  

This leads to some contradiction between the effectiveness and security of the existing quantum key distribution (QKD) schemes. The contradiction can be overcame by using multi-photon pulses and establishing the channel security on the realistic basis of multi-photon statistics rather then the fermion pairs statistics which the first QKD protocols where starting with.

Besides avoiding empty pulses by making carrier beams more intensive another point for increasing the effectiveness of quantum communications and QKD in particular is the alphabet size, which can be extended beyond the usual two-letter one,
which corresponds to a classical bit. It was shown that the large-alphabet coding essentially increases the bit rate of a quantum channel with no loss in security even in the case of a single-photon polarization coding \cite{larger}.

In this work we prove the effectiveness of the alphabet extension for the new multiphoton QKD protocol which was proposed recently \cite{tmcc,tmcc2} on the basis of the special beam states called two mode coherently correlated (TMCC). These states have the strong correlation between the photon numbers in each of the two spatially parted modes. This correlation leads to the fact that laser shot noise shows itself equally in the both of the modes thus enabling the use of a TMCC-source as a generator of some random bit sequence which is then shared between two legitimate parties who perform independent photon-number measurements on each of the modes and extract the key bits from the measurement results comparing them to the constant average photon number. The quantum channel capacity for the 4- and 8-letter extended alphabets is estimated. The eavesdropping attacks on the extended-alphabet TMCC-channels are considered in terms of the introduced QBER. It is shown that the alphabet extension leads to the  increase of the effectiveness of the QKD protocol providing channel capacity of up to 3 bits per pulse. Moreover the alphabet extension strengthen the protocol security against the intercept-resend attacks by making eavesdropper introduce larger QBER on each successfully intercepted bit exceeding 70\% QBER per bit for the 8-letter alphabet.
 
\section{Protocol}

The TMCC-protocol is based on the use of a special states of light, the 
two mode coherently correlated ones. Such states were mathematically studied by Agarwal \cite{agarwal1, agarwal2}. They are defined as fully correlated and at once the eigenstates for the product of annihilation operators of the both modes. The latter condition makes them the special case of the wider class of the two-mode correlated states, also referred to as twin beams, which are broadly investigated in the past time \cite{twin1,twin2,twin3} and are usually obtained in the process of parametric down-conversion (PDC) in nonlinear crystals.

The TMCC-states can be presented through series by Fock states:

\begin{equation}
\label{eq:tmcc}
\left| \lambda \right\rangle =\frac{1}{\sqrt{I_0\left(2\left|\lambda\right|\right)}}\sum_{n=0}^\infty {\frac{\lambda ^n}{n!}\left| {nn} 
\right\rangle } 
\end{equation}

Here we use the designation $\left| {nn} \right\rangle = \left| n 
\right\rangle _1 \otimes \left| n \right\rangle _2 $, where $\left| n 
\right\rangle _1 $ and $\left| n \right\rangle _2 $stand for the states of 
the $1^{st}$ and $2^{nd}$ mode accordingly, represented by their photon 
numbers.

The main feature of the states (\ref{eq:tmcc}) is that only the terms with equal photon numbers in the both of the modes are present in the expansion. This leads to the strong correlation between the observables concerned with each of the modes. At the same time the TMCC states differ from the usual two-mode correlated coherent states, because the average for any of the linear in field observable (e.g. the vector-potential) is equal to zero for each of the TMCC modes \cite{tmcc}. Thus each of the modes is not coherent to itself, but as the square in field observables (e.g. correlation function, momentum, energy) have non-zero average values, the TMCC states possess the second order coherence and so can be referred to as the coherently correlated states.

The separate intensity measurements on each of the beams give the results which are proportional to the average of the $N = a^ + a$, which corresponds to the number of photons in the mode. The probability distribution for different photon numbers detection is

\begin{equation}
\label{eq:nphotprob}
P_n(\lambda) = \frac{1}{I_0 (2|\lambda|)} \frac{|\lambda|^{2n}}{n!^2} 
\end{equation}

The photon statistics for each of the beams is sub-Poisson as confirmed by the Mandel parameter, which is negative even for the small intensities \cite{tmcc,tmcc2}. This fact can be useful for the TMCC state identification and the quantum channel eavesdropping detection. 

The shot noise photon number fluctuations in the TMCC modes enable the use of the TMCC source as a generator of a random key encoded in the photon numbers value. 
The strong correlation between the independent measurements of two modes makes it possible to securely distribute such random key between two remote legitimate users, Alice and Bob. The security of the quantum channel is based on the fact that intermediate photon measurements will perturb the states which can be checked locally by measurement statistics \cite{tmcc2}.

The simple protocol based on the two mode coherently-correlated states was described \cite{tmcc} as follows: Alice and Bob simultaneously start the independent photon number measurements each on the corresponding TMCC mode. They compare the obtained photon number values for each next unit time to the average which is constant during the overall key transmission procedure. If the obtained number is larger than the average, the next bit is considered to be equal to "1". If the photon number for the next time window is less than average, the corresponding bit is equal to "0". Thus, the protocol uses the two-bit photon number alphabet coding by the multiphoton two mode coherently correlated states.

The state coherence is an important feature of the TMCC-protocol distinguishing it from the existing single-photon incoherent state protocols because coherence allows establishing the security for the multiphoton pulses transmission since the state perturbation leads to the decoherence. The loss of coherence can be revealed by checking the state statistics which can be done even locally by estimating differences between the obtained and expected state density matrices \cite{tmcc2}.

\section{Alphabet extension}

The classical information channel is well known to be described by the Shannon mutual information between the observables of some classical macroscopic systems. The measurements on such systems return the probability distributions for the sets of the observables' discrete values. These value sets when used for encoding and decoding the information are called the alphabets which contain the discrete values of the observables as the letters. The input-output mutual information depends on the observables' Shannon entropies:

\begin{equation}
I(X;Y)=H(X)-H(X|Y),
\end{equation}

where $H(X)$ is the input observable entropy and the $H(X|Y)$ is the mutual entropy of the input relative to the output which describes the information loss in the channel:

\begin{equation}
H(X|Y) = - \sum_{x,y} p(x,y) \log p(x|y) = H(X,Y)-H(Y).
\end{equation}

Since $H(X|Y)$ depends on the probability distribution of the values set

\begin{equation}
H(X)  = - \sum_{x\in X} p(x) \log p(x),
\label{eq:shentropy}
\end{equation}

it depends on the alphabet size and increases with the increase of the number of its letters.

This discourse clearly fits the quantum channels which differ from the classical ones by the fact that quantum observables, being the parameters of the quantum microscopic states, are used for information encoding and decoding. Since quantum cryptography deals with the key sharing across the quantum channel it can also gain from the alphabet expansion. 

In 1999 Bechmann-Pasquinucci and Tittel \cite{larger} proposed the use of a larger alphabet for the BB84 single-qubit protocol \cite{bb84}, which is extended on the four-level quantum systems - the so-called quantum quarts (qu-quarts). Alice still selects randomly between the two possible bases, but she is now preparing one of the four states thus making eight possible choices for the qu-quart based protocol instead of the four choices for the qubit-based one. It was shown that such a development of the BB84 protocol increases the information flux and makes the QKD scheme more secure against realistic eavesdropping because eavesdropper introduces the much higher QBER for the given amount of the acquired information. 

Lately in 2003 Sych, Grishanin and Zadkov \cite{cont} made the further development of the single-qubit QKD scheme by proposing the use of the continuous alphabet for the key bits coding. The idea was to identify key bits from the unselected qubit states which was shown to result in the increase of the protocol effectiveness, security and reliability at noisy channels. 

Thus it makes it clear that the protocol extension may be quite useful for the TMCC-based protocol especially since the multi-photon states in the Fock presentation can be considered as the multi-dimensional systems. 

The maximum Shannon entropy of a photon-counting beam measurement, i.e. the highest possible information capacity for a channel built on such beams can obviously be achieved if each of the different photon-number events are identified as the different measurement outcomes and correspond to the different alphabet letters. In other words, the m-letter alphabet for the m-photon state will give the maximum information which can be encoded into and transmitted by such a state. For the TMCC-beam this maximal information will be

\begin{equation}
\label{eq:maxinfo}
H_{max}(\lambda) = \sum_{n=0}^m{\frac{1}{I_0 (2|\lambda|)}\frac{|\lambda|^{2n}}{n!^2} 
\log{(\frac{1}{I_0 (2|\lambda|)}\frac{|\lambda|^{2n}}{n!^2}})} 
\end{equation}

\begin{figure}[htbp]
\begin{center}
	\epsfbox{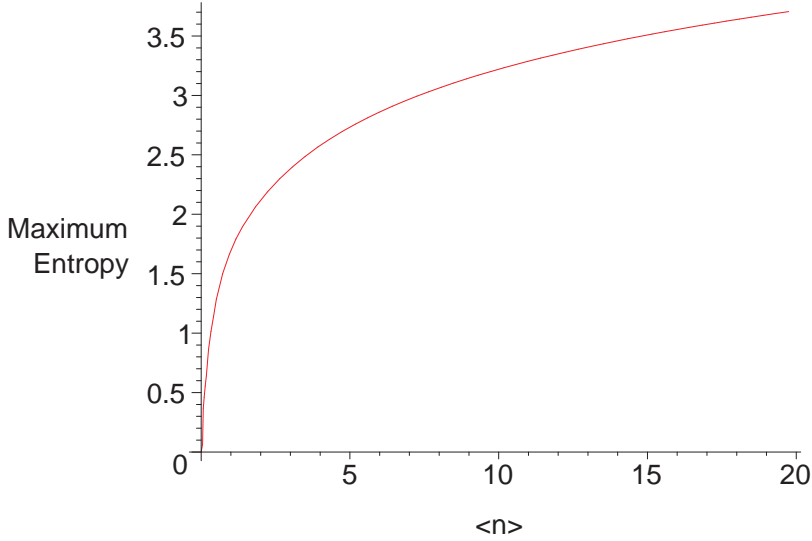}
	\caption{Maximum Shannon entropy of the photon-number measurements on the TMCC-states}
	\label{maxent}
\end{center}
\end{figure}

The dependence of this highest possible information gain of a state measurement on the average photon number of the state is given at (\ref{maxent}).

One can easily see that depending on the beam intensity, the TMCC state can carry from 1 up to 4 bits of information. Thus it is possible to build the effective TMCC-based QKD protocols utilizing 4-letter alphabets for beams carrying about 3-5 photons in average which will raise the measurement information gain to 2 bits and the 8-letter alphabets for more intensive beams which will result in the effectiveness of up to 4 bits per measurement.

Here we examine two possible alphabet extensions for the TMCC channels, containing 4 and 8 state "letters".

\begin{figure}[htbp]
\begin{center}
	\epsfxsize=400pt
	\epsfbox{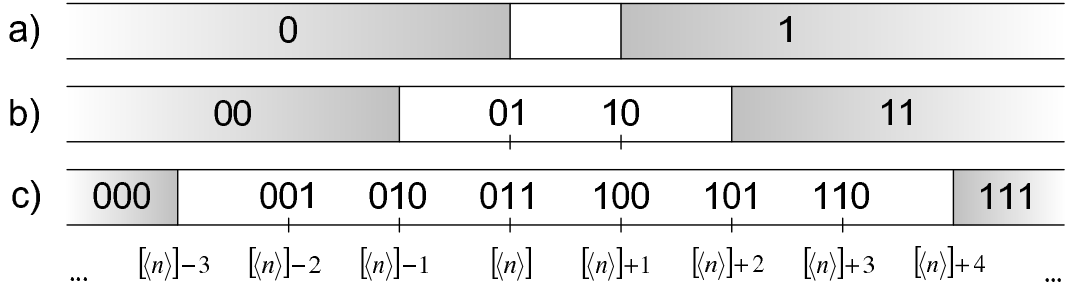}
	\caption{Graphical representation of three alphabet sets for a channel based on the multiphoton pulses: a) usual 2-letter alphabet with the capacity of 1 bit per photon-number measurement; b) 4-letter alphabet with 2 bits available from each measurement and c) 8-letter alphabet with 3 bits information gain for each measurement.}
	\label{bits}
\end{center}
\end{figure}

The quaternary channel with two bits per measurement capacity can be established on the basis of the TMCC beams which are distinguished by the photon numbers between 4 possible states (see figure \ref{bits}):

\begin{equation}
\label{eq:quart}
Q ={\left\{
\begin{array}{l@{\}\rightarrow}l}
\{n \leq \nav-1 & 00$ (letter 0)$\\ 
\{n = \nav & 01$ (letter 1)$\\ 
\{n = \nav+1& 10$ (letter 2)$\\ 
\{n \geq \nav+2& 11$ (letter 3)$
\end{array}\right.}
\end{equation}

Knowing the photon numbers registration probability distribution for a TMCC-beam (\ref{eq:nphotprob}) one can easily estimate the probability for each of the 4 quart states (\ref{eq:quart}) realization:

\[P_0(\lambda) = \frac{1}{I_0 (2|\lambda|)} \sum_{n=0}^{\nav-1}{ \frac{|\lambda|^{2n}}{n!^2} };\ P_1(\lambda) = \frac{1}{I_0 (2|\lambda|)}  \frac{|\lambda|^{2\nav}}{\nav!^2};
\]
\begin{equation}
P_2(\lambda) = \frac{1}{I_0 (2|\lambda|)}  \frac{|\lambda|^{2(\nav+1)}}{(\nav+1)!^2};\ P_3(\lambda) = \frac{1}{I_0 (2|\lambda|)} \sum_{n=\nav+2}^\infty{ \frac{|\lambda|^{2n}}{n!^2} }
\end{equation}

These probability values can be then used for estimating the Shannon entropy (\ref{eq:shentropy}) of a photon-number measurement on a TMCC-beam with four possible interpretation outcomes, which is the information gain for such a measurement.

The octuple TMCC-based channel with the capacity of three bits per measurement can be established in a similar way if the states are distinguished by 8 possible measurement outcomes which differ in photon numbers (see figure \ref{bits}):

\begin{equation}
\label{eq:oct}
O ={\left\{
\begin{array}{l@{\}\rightarrow}l}
\{n \leq \nav-3& 000$ (letter 0)$ \\ 
\{n = \nav-2& 001$ (letter 1)$ \\ 
\{n = \nav-1& 010$ (letter 2)$ \\ 
\{n = \nav & 011$ (letter 3)$ \\ 
\{n = \nav+1& 100$ (letter 4)$ \\ 
\{n = \nav+2& 101$ (letter 5)$ \\ 
\{n = \nav+3& 110$ (letter 6)$ \\ 
\{n \geq \nav+4& 111$ (letter 7)$
\end{array}\right.}
\end{equation}

Again, knowing the distribution (\ref{eq:nphotprob}) we can obtain the probability for each octo-bit value occurrence and thus build the Shannon information available from a measurement.

\begin{figure}[htbp]
\begin{center}
	\epsfbox{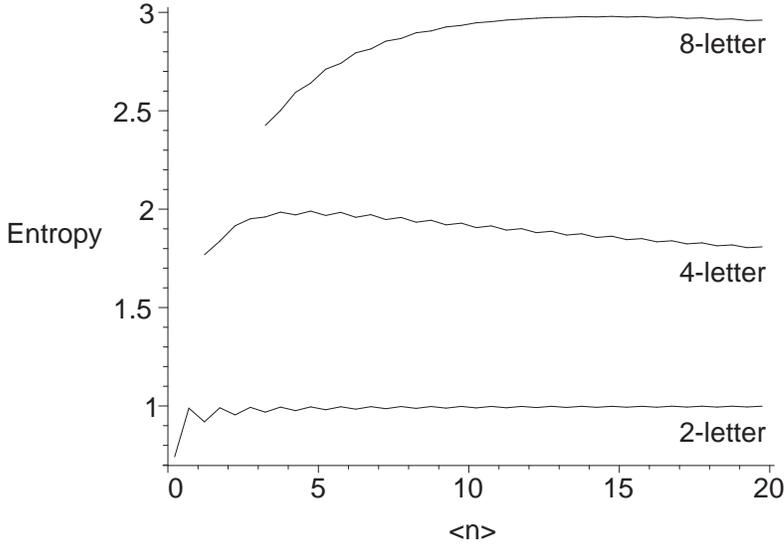}
	\caption{Shannon entropies characterizing information gain of the TMCC-states photon-number measurements for binary, quaternary and octuple alphabets}
	\label{entropies}
\end{center}
\end{figure}

The dependencies for the measurement information gain for 2-, 4- and 8-letter alphabets on the state average photon number are given on a graph at (\ref{entropies}).

\section{Security}
The security of a quantum channel i.e. the impossibility to carry out a successful eavesdropping is obviously the most important property for any QKD scheme. Traditionally the security of a quantum channel is examined in the frames of the two approaches - the ideal and the realistic ones \cite{lutkenhaus2}. In the ideal case the eavesdropper Eve is supposed to have the unlimited technological power with possibilities restricted only by the laws of quantum mechanics \cite{lo}. The realistic approach takes into account the technical possibilities of an eavesdropper compatible with today's and foreseeable technology at this the most successful realistic eavesdropping techniques on the secure quantum channels are the intercept-resend strategy also referred to as state cloning and the beam splitting \cite{felix}. The most general security criterion known from the classical cryptography already states that Bob has to possess more information on the transferred key than Eve (if the Bob's mode was eavesdropped). In this case the privacy amplification post-transfer algorithms will be helpful in distilling the truly secure key, otherwise they will fail \cite{main}. 

The eavesdropping in quantum cryptography is usually detected by checking the QBER (quantum bit error rate) which is the measure of the errors in the obtained key. Since all of the errors are considered to be caused by an eavesdropper, knowing the QBER one can estimate how much information on the key does the eavesdropper have and thus determine is the key distillation will be successful. 

The TMCC-based scheme was examined against the realistic eavesdropping and shown to be secure against splitting and cloning attacks. In the case of the beam splitting this is quite intuitive because the installation of a splitter at any of the modes removes the correlation between modes thus simply destroying the channel (for more detailed examination of a beam splitting attack on a TMCC-channel see \cite{tmcc,tmcc2}). In the case of a state cloning it was shown that eavesdropper significantly changes the statistics of the re-emitted mode which can be detected by the local calculations of the Mandel parameter or the distances between the received and the expected states' density matrices \cite{tmcc2}. In this work we estimate the security of the TMCC-based channel in the terms of the QBER, which is introduced to the channel upon the eavesdropping. We use this measure to compare the security of the TMCC QKD scheme for different alphabet sizes.

\begin{figure}[htbp]
\begin{center}
	\epsfbox{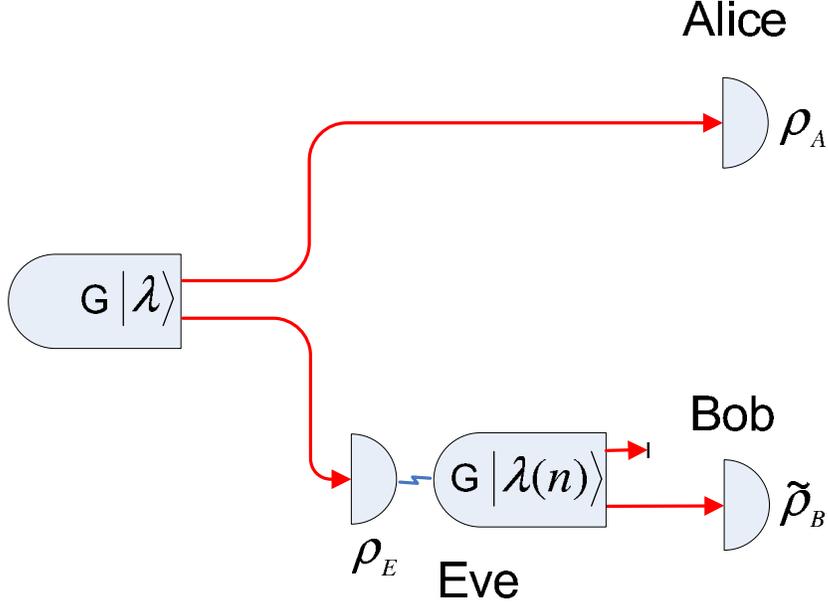}
	\caption{State cloning intercept-resend eavesdropping attack on a TMCC-based channel}
	\label{cloning}
\end{center}
\end{figure}

Let's consider Eve carrying out a cloning intercept-resend attack on a TMCC-based channel. In order to do so she installs a photon-counting detector which measures one of the modes (suppose the one which goes to Bob) and a TMCC laser source assigned for re-creating the state (fig \ref{cloning}). 

For each incoming pulse (i.e. in each time slot) Eve measures the photon number $n$ and tries to re-emit the same photon number in the Bob's direction. We assume Eve calculates the value of state parameter $\lambda_n$ which corresponds to the photon number $n$ to be emitted and sets her source up to this value. Though due to the quantum fluctuations she is unable to emit exactly the same number and thus Bob obtains the state which density matrix $\tilde {\rho }_B$, differs from the original $\rho_B$ measured by Eve. This re-emitted state matrix is the mixture of the k-photon states 

\begin{equation}
\tilde {\rho }_B=\sum\limits_{k = 0}^\infty {\tilde {P}_k 
\left| k \right\rangle \otimes \left\langle k \right|}
\end{equation}

with probabilities 
\begin{equation}
\tilde {P}_k = 
\sum\limits_{n = 0}^\infty {
\frac{\left| \lambda \right|^{2n}}{n!^2I_0 
(2\left| \lambda \right|)}
\frac{\left| {\lambda_n} \right|^{2k}}
{k!^2I_0(2\left| {\lambda_n} \right|)}} .
\end{equation}

The probability for Bob to incorrectly interpret the received state i.e. to obtain the wrong letter $x$ of the alphabet set $X$ of size $m$ (so that $X=\{x_0...x_m\}$) is equal to

\begin{equation}
P_{err}=\sum_{x \in X}{\frac{1}{m}(1-P_B(x|x))},
\label{eq:perr}
\end{equation}

where $P_B(x|x)$ is the probability for Bob to obtain the correct letter $x$ given Alice obtained the same letter $x$. The QBER introduced by Eve for each intercepted bit is then

\begin{equation}
P_{err/bit}=\frac{1}{\log_2{m}}P_{err}.
\end{equation}

The sum in $\ref{eq:perr}$ can be presented and calculated through the probability distributions as

\begin{equation}
\sum_{x \in X}P_B(x|x)=\sum_{k,n=0}^{n_0}{P_k(\lambda_n)P_n(\lambda)}+
\sum_{k,n=n_m}^{\infty}{P_k(\lambda_n)P_n(\lambda)}+
\sum_{n=n_1}^{n_m-1}{P_n(\lambda)},
\end{equation}

where $n_0$ is the photon number corresponding to the first alphabet letter $x_0$ and $n_m$ is the photon number corresponding to the last letter $x_m$, $P_n(\lambda)$ is the probability distribution ($\ref{eq:nphotprob}$) and 

\begin{equation}
P_k(\lambda_n)=\frac{\left| {\lambda_n} \right|^{2k}}
{k!^2I_0(2\left| {\lambda_n} \right|)}
\end{equation}

\begin{figure}[htbp]
\begin{center}
	\epsfbox{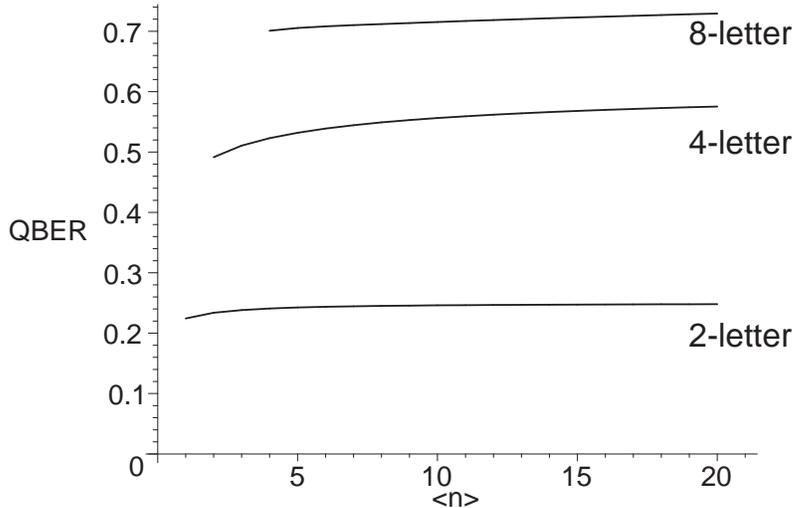}
	\caption{QBER introduced by Eve for each intercepted bit during the state cloning attack with a TMCC-source on a TMCC-based channel for different alphabet sizes dependence on the average photon number}
	\label{qber}
\end{center}
\end{figure}

Hence the QBER can be calculated numerically. The results are presented on the graphs at fig $\ref{qber}$ as the QBER dependence on the average photon number (i.e. on the original pulse intensity) for different alphabet sizes. One can easily see that QBER introduced for each intercepted bit during the state cloning attack on a 2-letter (1-bit) alphabet channel is about 20\%, growing to about 50\% in the case of 4-letter (2-bit) case and exceeding 70\% for the 8-letter (3-bit) alphabet. Thus the TMCC-based QKD protocol turns out to be more secure at the larger alphabets.

If Eve uses a usual single mode laser source, producing a coherent beam with the Poisson statistics for the TMCC-state cloning, the calculations give QBER values of about 30\% for 2-letter, up to 60\% for 4-letter and over 80\% for the 8-letter alphabet thus proving that TMCC-state cloning based on a usual laser source is even less effective.

\section{Conclusions}
The quantum cryptography scheme with larger alphabets based on the use of the two-mode coherently correlated multiphoton beams is proposed. The alphabet extension is shown to result in the increase of the effectiveness of the QKD scheme. The protocol security against the realistic state cloning is examined in terms of the introduced QBER. It is shown that the TMCC-based QKD scheme becomes more secure for the larger alphabet sets i.e. for the more intense laser pulses.

\end{document}